\documentclass{aastex631}%

\newcommand{\A}[1]{\textcolor{purple}{[AA: #1]}}

\usepackage{bbm}
\usepackage{enumerate}

\begin{document}

\title{Detecting Model Misspecification in Cosmology with Scale-Dependent Normalizing Flows}

\correspondingauthor{Aizhan Akhmetzhanova}
\email{aakhmetzhanova@g.harvard.edu}

\author{Aizhan Akhmetzhanova}
\affiliation{Department of Physics, Harvard University, 17 Oxford Street, Cambridge, MA 02138, USA}
\affiliation{The NSF AI Institute for Artificial Intelligence and Fundamental Interactions}

\author{Carolina Cuesta-Lazaro}
\affiliation{Harvard-Smithsonian Center for Astrophysics, 60 Garden Street, Cambridge, MA 02138, USA}
\affiliation{The NSF AI Institute for Artificial Intelligence and Fundamental Interactions}
\affiliation{Department of Physics, Massachusetts Institute of Technology, Cambridge, MA 02139, USA}

\author{Siddharth Mishra-Sharma}
\altaffiliation{Currently at Anthropic; work performed while at MIT/IAIFI.}
\affiliation{Department of Physics, Massachusetts Institute of Technology, Cambridge, MA 02139, USA}
\affiliation{The NSF AI Institute for Artificial Intelligence and Fundamental Interactions}
\affiliation{Center for Theoretical Physics, Massachusetts Institute of Technology, Cambridge, MA 02139, USA}
\affiliation{Department of Physics, Harvard University, 17 Oxford Street, Cambridge, MA 02138, USA}

\begin{abstract}
Current and upcoming cosmological surveys will produce unprecedented amounts of high-dimensional data, which require complex high-fidelity forward simulations to accurately model both physical processes and systematic effects which describe the data generation process. However, validating whether our theoretical models accurately describe the observed datasets remains a fundamental challenge. An additional complexity to this task comes from choosing appropriate representations of the data which retain all the relevant cosmological information, while reducing the dimensionality of the original dataset. 
In this work we present a novel framework combining scale-dependent neural summary statistics with normalizing flows to detect model misspecification in cosmological simulations through Bayesian evidence estimation. By conditioning our neural network models for data compression and evidence estimation on the smoothing scale, we systematically identify where theoretical models break down in a data-driven manner. We demonstrate a first application to our approach using matter and gas density fields from three CAMELS simulation suites with different subgrid physics implementations. 
\end{abstract}


\section{Introduction} \label{sec:intro}

Observational cosmology now faces a growing number of tensions that challenge the standard $\Lambda$CDM model, such as the persistent discrepancy in Hubble constant ($H_0$) measurements between early and late Universe probes \citep{Riess_2022} or the $S_8$ tensions regarding matter clustering strength \citep{cosmo_tensions_abdalla2022cosmology}. Moreover, the DESI collaboration has presented some preliminary evidence towards a preference for evolving dark energy from Baryon Acoustic Oscillation measurements \citep{Adame_2025}. These anomalies have so far been found by either \emph{(i)} direct comparison of parameter constraints derived from different cosmic epochs, where agreement would validate our understanding of cosmological evolution, or \emph{(ii)} parametric extensions to $\Lambda$CDM, such as the $\{w_0,w_a\}$ dark energy parameterization, which test specific model extensions beyond $\Lambda$CDM. In this paper, we present a complementary approach focused on identifying model-independent anomalies. That is, once the baseline model or training dataset has been specified, our approach for detecting anomalous data does not require that the anomalies follow a particular functional form or parameterization, or even a parameter-free ansatz. 

We take advantage of advances in the availability of large datasets of numerical simulations \citep{Pakmor_2023, Maksimova_2021, Villaescusa_Navarro_2021}, precise observations spanning large volumes from surveys such as DESI and Euclid, and high-dimensional inference techniques capable of solving complex inverse problems.  Although most of the research at this intersection has focused on parameter estimation and optimal information extraction, we propose a shift in perspective to tackle a complementary question: \emph{how can we systematically identify significant discrepancies between our theoretical models, as represented by numerical simulations, and the observed Universe?}

Finding discrepancies between simulations and observations serves a second important purpose beyond anomaly detection: it addresses the critical need for robust goodness-of-fit metrics in high-dimensional spaces to validate our inference methods. Recent years have seen inference models primarily focused on parameter estimation without adequate assessment of model fit. Traditional approaches like $\chi^2$-statistics are insufficient for this task due to their limitations with high-dimensional data, such as having access to a large number of samples to compute these statistics reliably), and the assumptions about about Gaussianity of the errors and linearity of the models. 
This concern is particularly relevant in cosmology, where potential model misspecification arising from theoretical uncertainties in galaxy formation or unaccounted survey systematics can significantly bias our inferences when incorporating small-scale information.

Detecting out-of-distribution (OOD) data is crucial to ensuring that trained machine learning systems are applied in a safe and reliable manner, given that small shifts in the data distribution can introduce large biases in the parameters of interest, see \citep{horowitz2022plausibleadversarialattacksdirect,nayantara_mudur2024diffusion} for examples from cosmology. To this end, the topic of OOD detection has attracted increasing interest from the machine learning research community, with multiple new methodologies proposed to address different aspects of the OOD detection problem, including detection of anomalies in test data. Broadly, these methods can be divided into four distinct categories: classification-based methods, density-based methods, distance-based methods, and reconstruction-based methods. A detailed exploration of these approaches to OOD detection can be found in \cite{yang2024_generalized_OOD_survey}. 
Of particular interest to our problem are density-based methods, which probabilistically model the distribution of the training (in-distribution) data. These methods then detect anomalies or OOD examples in the test data by flagging data points which lie in the low-density regions of the probability space. Density-based approaches greatly benefit from advancements in deep generative models, such as normalizing flows~\citep{papamakarios2021normalizing} and diffusion models~\citep{diffusion_sohl2015deep, diffusion_song2020score}, which can estimate complex density distributions, often from high-dimensional inputs, and provide access to the exact marginal likelihood of the data. 

Our work builds on recent advances in neural density estimation and anomaly detection to develop a scale-dependent framework for identifying model misspecification in cosmological analyses. Physical models in cosmology often have well-defined domains of validity that depend on spatial scale; for instance, our understanding of large-scale structure may be accurate on larger scales where gravity dominates, but less reliable on smaller scales where baryonic physics becomes important. By explicitly incorporating scale dependence into our analysis, we can identify at which scales theoretical models begin to break down.

Previously, \cite{dai2024multiscale} explored the idea of identifying anomalies and distribution shifts due to scale-dependent systematics at the field-level with Multiscale Flows, which hierarchically decompose cosmological fields into lower-resolution approximations using a wavelet basis and use normalizing flows to model each wavelet component separately and learn the full field-level likelihood function. They applied Multiscale Flows to simulated weak lensing datasets and found that in addition to significantly improving cosmological inference analysis, these flows are also able to detect small-scale OOD shifts such as addition of baryonic effects. Here, we develop an alternative framework for multiscale neural posterior estimation such that we can scale to higher dimensional datasets.

Concurrent to our work, \cite{diao2025detecting} has extended \cite{dai2024multiscale} by approaching the task of model validation as an OOD detection problem. Similarly to \cite{dai2024multiscale}, they work with simulated weak lensing datasets generated from dark-matter only simulations and corrections from baryonic effects. They examine multiple various statistics on these datasets, and find that continuous time flow models at the field level consistently achieve the best performance for their OOD detection task. 

These studies highlight the relevance of density-based OOD detection methods in cosmology in the context of model validation and anomaly detection. Our work complements the studies above and extends this line of work in new directions. In particular, our contributions are: \emph{(1)} we develop a general training strategy for obtaining constraints as a function of scale on one single model, \emph{(2)} we demonstrate that estimating evidence as a function of scale on learned summary statistics which are optimized for cosmological parameter inference can be used for detecting model misspecification through an example using the CAMELS simulation suite.

The paper is structured as follows. We present our methodology in Section \ref{sec:methodology}, along with the implementation details, training procedure, and architectures of the neural networks used in this study. In Section \ref{sec:results} we describe the experimental setup employed to validate our framework and examine its performance in a controlled setting. Finally, we conclude with a summary of our results and a discussion of potential extensions of this framework to real cosmological survey data in Section \ref{sec:summary}.

\section{Methodology} \label{sec:methodology}

In this paper, we use the Bayesian evidence as a natural out-of-distribution score by comparing observed and simulated samples. 
We first learn lower-dimensional representations of our high-dimensional observables that are optimized to constrain cosmological and astrophysical parameters of interest. 
Starting from these learned summary statistics, we use the Bayesian evidence -- the probability of an observation under our simulations marginalized over all simulation parameters -- to assess model misspecification. 

\begin{figure}[t!]
    \centering
    \includegraphics[width=\linewidth]{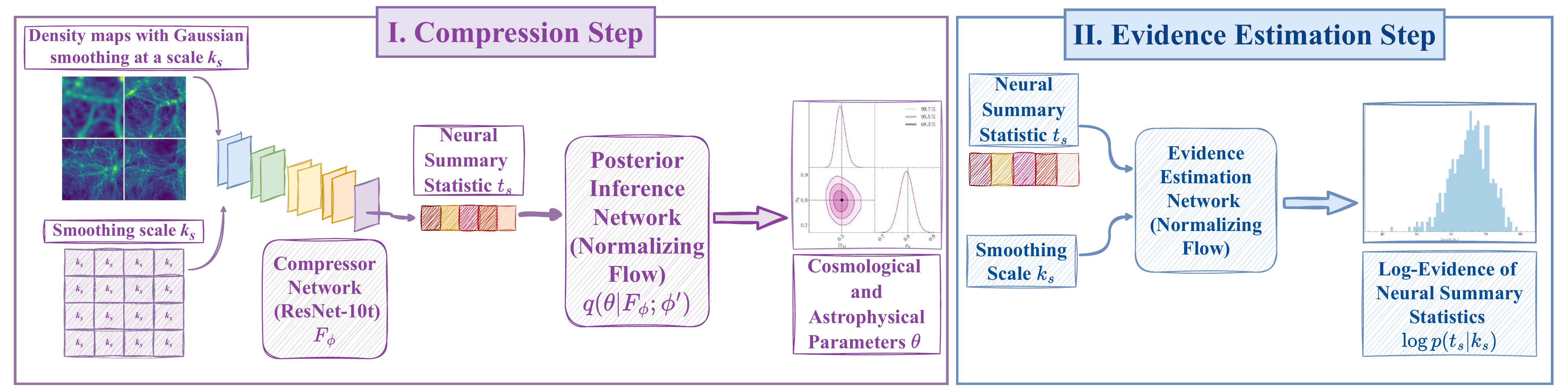}
    \caption{A schematic overview of the framework for OOD detection implemented in this study. As is shown on the left panel, we first learn sufficient neural summary statistics of the data (e.g. total matter and gas density fields) by training a compressor network with a Variational Mutual Information Maximization (VMIM) loss. The right panel illustrates the second step of our framework in which we train a normalizing flow to estimate the evidence of the learnt summary statistics as a function of smoothing scale $k_s$. }
    \label{fig:pipeline}
\end{figure}

Of particular importance in physics is how out-of-distribution metrics vary as a function of physical scale. 
This is especially relevant given that we often have accurate descriptions of large-scale phenomena through effective field theories~\citep{eft_cosmo_cabass2023snowmass}, while missing crucial details at smaller scales with higher frequency components. 
In cosmology, this manifests as theoretical uncertainties about how small-scale processes -- such as supernovae and AGN feedback—impact the larger-scale distribution of matter~\citep{chisari2019modelling}. 
To address this specific challenge, we present a tailored solution that conditions our neural summary statistics on the smoothing scale, allowing us to systematically analyze how model misspecification varies across different physical scales. 

\begin{figure}[ht!]
\plotone{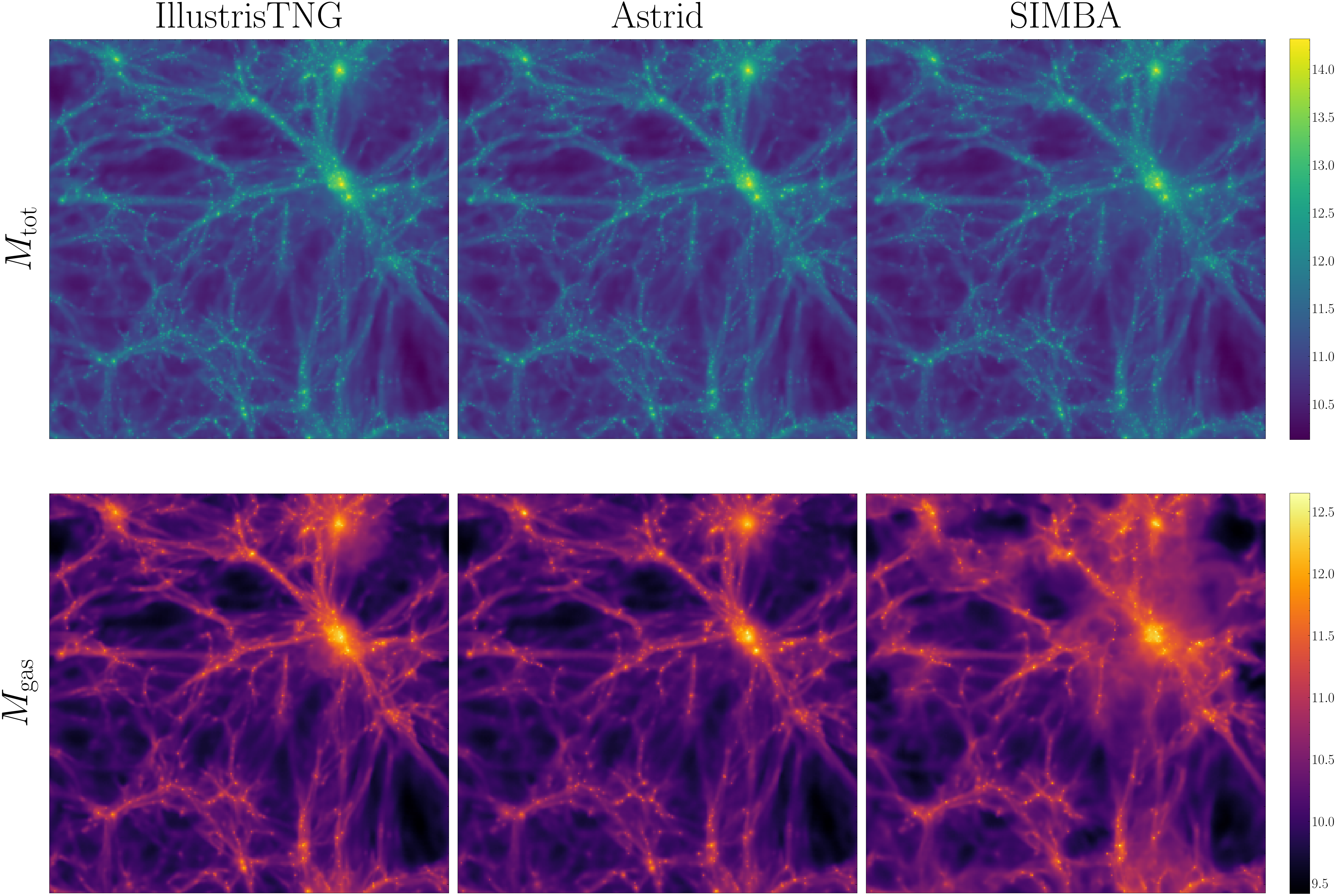}
\caption{Examples of projected total matter density $M_{\mathrm{tot}}$ (top row) and gas density $M_{\mathrm{gas}}$ (bottom row) fields from the CV sets of IllustrisTNG (left), Astrid (middle), and SIMBA (right) simulations suites. For visual comparison of the three subgrid physics models, the examples are taken from the simulation boxes which share the same random seed for initial conditions. The maps are plotted on the log scale (with base 10). }\label{fig:astro_fields}
\end{figure}

Figure~\ref{fig:pipeline} provides a schematic overview of the framework proposed in this study. In Section~\ref{sec:results} we demonstrate an application of our methodology to the problem of cosmological parameter estimation from observations of the Universe's large scale structure. In particular, we use the CAMELS suite of hydrodynamical simulations \citep{CAMELS, CAMELS_CMD, CAMELS_Astrid} to demonstrate that our models can detect the differences in the subgrid physics and how the detection strength varies as a function of scale and observable. In Figure~\ref{fig:astro_fields}, for visual comparison, we showcase two different fields used in this study, total matter $M_\mathrm{tot}$ and gas mass $M_\mathrm{gas}$ distribution, for three different subgrid models (IllustrisTNG, Astrid, and SIMBA) from the CAMELS suite. 

\subsection{Learning Scale-Dependent Summary Statistics}\label{subsec:neural_summary_npe}
When learning neural representations of our observables, we aim to obtain sufficient statistics from high-dimensional data that capture all relevant information about the parameters of interest. 
A statistic $t$ is sufficient for parameters $\theta$ if and only if $I(x, \theta) = I(t, \theta)$, such that $t$ preserves all information from the original data $x$ that is relevant for inferring $\theta$. 
We follow the neural posterior estimation (NPE) approach to learn these summary statistics \citep{papamakarios2018fastepsilonfreeinferencesimulation}. 
The summary features $t = F_\phi(x)$ are extracted from the data with a compressor network $F_\phi$. The joint posterior distribution of the parameters interest is modeled with a normalizing flow $q_{\varphi}(\theta|t)$ (with $\varphi$ being the parameters of the flow) \citep{rezende2015variational,papamakarios2021normalizing} which is conditioned on the summary statistics $t$. 
The flow transformation and the compressor network are jointly trained by minimizing the negative log-posterior density:
\begin{equation} \label{eq:npe_loss_function}
\mathcal{L} = -\log q_{\varphi}(\theta|F_\phi(x)).
\end{equation}

This approach (also referred to as Variational Mutual Information Maximization (VMIM) \citep{jeffrey2021likelihood,lanzieri2024optimal,Ho_2024}) works by maximizing the mutual information $I(t, \theta)$ between cosmological parameters $\theta$ and summary statistics $t$: $I(t, \theta) \equiv \mathbb{E}_{p(t,\theta)}[\log p(\theta|t)] - H(\theta)$. Since the entropy $H(\theta)$ is constant with respect to the compressor network, maximizing $I(t, \theta)$ is equivalent to maximizing $\mathbb{E}_{p(t,\theta)}[\log p(\theta|t)]$. However, the true posterior $p(\theta|t)$ is unknown, so with NPE we instead optimize a variational lower bound on $I(t, \theta)$:

\begin{equation} \label{eq:mi_lower_bound}
I(t, \theta) \geq \mathbb{E}_{p(t,\theta)}[\log q_\varphi(\theta|t)] - H(\theta).
\end{equation}

With sufficient network capacity and training data, following this approach allows one to construct summary statistics that are theoretically optimal for parameter inference, ensuring minimal information loss during compression. 

\subsubsection{Architecture of the compressor network and the posterior inference network} \label{subsubsec:compressor_architecture}
The compressor network takes as input $256 \times 256$ simulated density fields and outputs summary statistics vector $t$. 
For the compressor network, we use a ResNet-10-T model as implemented in \cite{timm_repository}\footnote{https://github.com/huggingface/pytorch-image-models}. 
This model is a lightweight variation of the ResNet architecture \citep{ResNet_he2016}, with fewer trainable parameters (4.94 million parameters), which we found to perform better on our dataset than larger models, such as ResNet-18 (11.2 million parameters)  or ResNet-34 (21.3 million parameters). 
The model consists of four residual blocks, with one convolutional layer in each block, followed by average pooling and downsampling. 
Throughout the network, we use circular convolutions to account for the periodic boundary conditions of the input fields. 
The final output of the network is a summary statistics vector $t$, the dimensionality of which we fix to $d=40$. 

The compressor network is trained jointly with a posterior inference network $q(\theta|t; \phi')$ to learn the optimal summary statistics. The posterior inference network is a Masked Autoregressive Flow (MAF), which is conditioned on the summary statistic $t$ from the compressor network, and is trained to predict a vector of all 6 cosmological and astrophysical parameters. The MAF consist of 8 transforms. Each transform a masked MLP (multi-layer perceptron) composed of 3 layers with 256 hidden features in each layer and ReLU activation functions between the layers. To help facilitate the training, each parameter is normalized with respect to its range such that its value lies between [-1, 1]. 

\subsubsection{Incorporating scale conditioning}\label{subsubsec:including_scale_cond}

We condition the compressor network on the smoothing scale such that we can evaluate the sufficient summary statistics $t$ at any scale $k_s$ with the same set of neural network parameters, $\phi$:

\begin{equation} \label{eq:summary_stat_definition}
 t_s = F_\phi(x_s),
\end{equation}

where $x_s$ is a Gaussian smoothed field, with smoothing scale $k_s$.

We incorporate scale conditioning by passing the smoothing scale $k_s$ as an additional channel to the compressor network, prior to the first convolutional layer. 
With $k_s$ sampled logarithmically, we find that taking $\log_{10}k_{s}$ and normalizing the resulting value by the logarithm of the smallest smoothing scale, $\log_{10}k_{s, \mathrm{max}}$, improves the performance of the compressor network in terms of the parameter constraints, compared to using the raw values of $k_s$. 
We investigated a few other approaches for conditioning on the smoothing scale $k_s$, such as including $k_s$ as an input to the final fully-connected layer of the compressor network, both as a single value and as an embedding in a higher-dimensional vector space, and adding an embedding of the smoothing scale to the output of the first convolutional layer of the compressor network. However, we find that the simple scale conditioning method described above either outperforms or results in comparable performance to these approaches, so we choose to use this implementation of scale conditioning in our study.

We train the compressor network for smoothing scales ranging from $k_{s, \mathrm{min}} = 2 \, h/ \mathrm{Mpc}$ to $k_{s, \mathrm{max}} = 45 \, h/ \mathrm{Mpc}$, which encompass a wide range of scenarios: at the lower end of the $k_s$ range (large smoothing scales) only the large-scale features are recovered in the fields, while the fields with large $k_s$ values (small smoothing scales) have resolution that is comparable to the resolution of the original fields. Figure~\ref{fig:npe_constraints_scale} provides visual examples of the effects of Gaussian smoothing on scales from this range. 

During the training, for each batch we sample the smoothing scales logarithmically, and apply the corresponding Gaussian smoothing to the fields on the fly. This flexible sampling approach makes it unlikely that the model encounters the same data sample (i.e. an augmented field with the exact same smoothing scale applied to it) more than once, which encourages the model to generalize better to the unseen data.  

\subsection{Evidence estimation with Normalizing Flows}\label{subsec:evidence}

\begin{figure}[t!]
    \centering
    \includegraphics[width=0.8\linewidth]{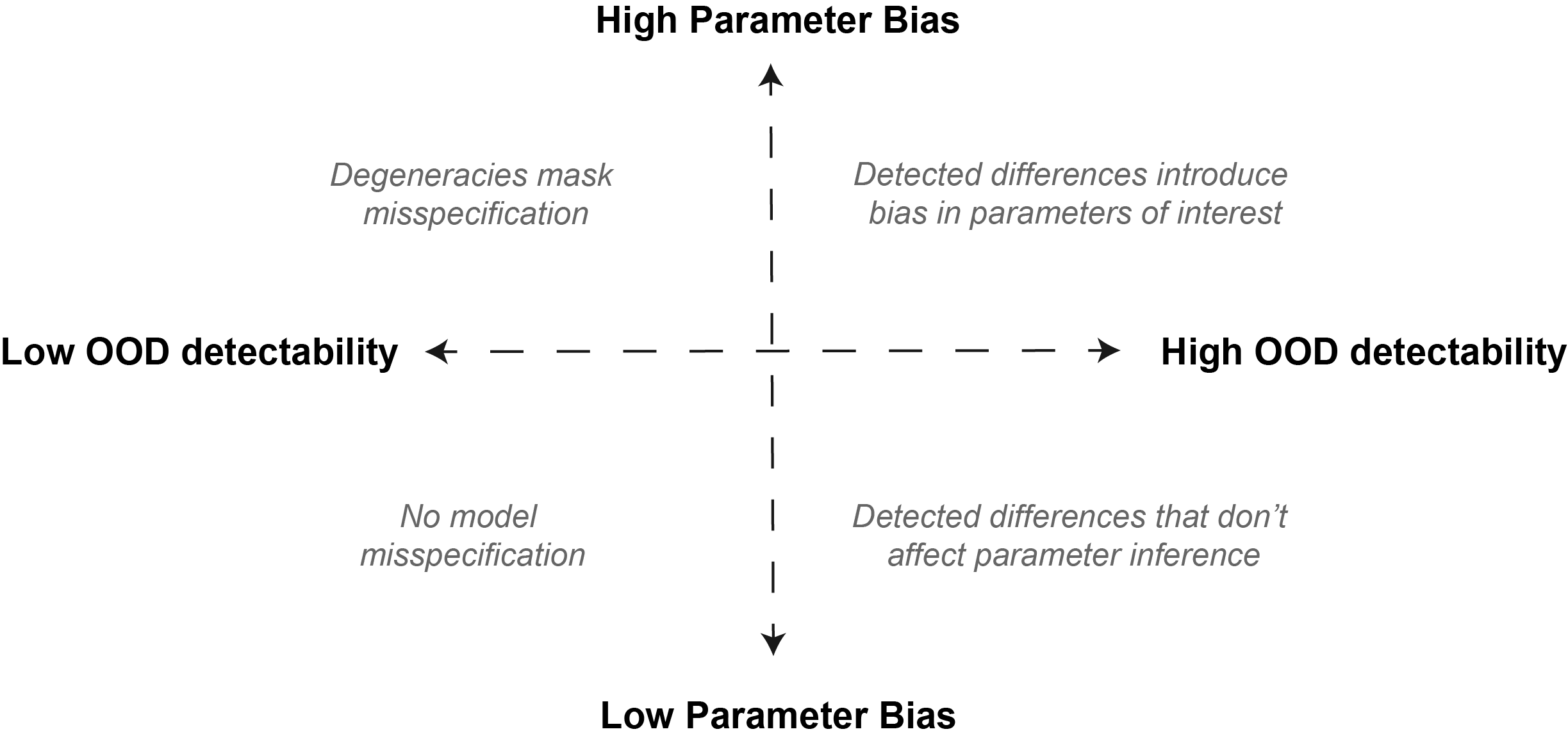}
    \caption{Interpretations of possible scenarios of out-of-distribution (OOD) detectability and biases on inferred parameters.}
    \label{fig:quadrant}
\end{figure}

Bayesian evidence (also known as the marginal likelihood) is a crucial quantity in model comparison and anomaly detection. For a model with parameters $\theta$ and data $x$, the evidence is defined as the probability of the data $x$ under the given model:

\begin{equation} \label{eq:evidence_definition}
p(x) = \int p(x|\theta)p(\theta)d\theta,
\end{equation}

which is computed by marginalizing the likelihood of the data $p(x|\theta)$ over the parameter space. This integral is often intractable, especially for high-dimensional parameter spaces or complex likelihoods. Various machine learning approaches have been developed to address this challenge~\citep{skilling2004_evidence, papamakarios2021normalizing, jeffrey2024_evidence, polanska2024_learned_hme_evidence}. These include direct density estimation via normalizing flows and variational methods that provide lower bounds on the evidence. Each method presents different trade-offs between accuracy, computational efficiency, and ease of implementation.

In this work, we focus on density estimation using normalizing flows applied to the low-dimensional latent space of our neural summary statistics. Specifically, after learning compressed representations $t_s = F_\phi(x_s)$ that are optimized for parameter inference, we train normalizing flows to model the density $p(t_s)$ in this lower-dimensional space, also as a function of smoothing scale. The reduction in dimensionality makes density estimation significantly more tractable and avoids many of the challenges typically associated with density estimation for OOD detection \citep{nalisnick2018_OOD, zhang2021_understanding_failures_OOD}, while our VMIM-optimized summaries retain the key information needed for parameter inference. We refer to the the normalizing flow trained to estimate the density of the compressed representations $p(t_s)$ as the \textit{evidence estimation network}.

For anomaly detection and model misspecification testing, we leverage the evidence as a natural out-of-distribution score by comparing it to the evidence of the in-distribution simulation samples. When a new observation yields summary statistics with unusually low probability density under our trained flow model, this indicates potential model misspecification or missing physics in our simulations. Detecting misspecification in raw data would not necessarily lead to this conclusion, as anomalies in data space might reflect physics irrelevant to our cosmological constraints, such as uncorrelated effects that differ from simulations but do not impact our parameters of interest. By examining how these evidence values vary across different physical scales, we can identify specific scales where discrepancies between our simulations and observations are most prominent, potentially revealing limitations in our physical models. This scale-dependent analysis provides a powerful diagnostic tool that goes beyond simple accept/reject decisions, offering insight into where and how our models might need refinement.

However, it is important to note a fundamental limitation of \textit{any} OOD detection method in this context: even when observations appear in-distribution according to our metrics, we may still obtain biased parameter constraints. This scenario arises when different physical effects produce features in our summary statistics that are statistically indistinguishable from one another, meaning that they are degenerate. For instance, certain baryonic feedback processes might mimic the signal of varying a particular cosmological parameter, like $\sigma_8$. In such cases, our neural summary statistics may fail to capture the distinction because the features map to similar signals in our compressed representation space. Consequently, we may incorrectly infer that our simulations adequately describe the observed universe when, in fact, they miss key physics. This limitation will be present in any OOD detection method, as it reflects a fundamental degeneracy in the observable quantities rather than a failure of the algorithm itself. The most robust approach to addressing such degeneracies would involve combining multiple complementary probes that respond differently to various physical effects, thereby breaking these degeneracies in the joint analysis. We have summarized the four possible relations between OOD detectability and parameter biases in Figure~\ref{fig:quadrant}.

\subsubsection{Architecture of the evidence estimation network}\label{subsubsec:evidence_network_architecture}

We refer to the the normalizing flow trained to estimate the density of the compressed representations $p(t_s)$ as the evidence estimation network. 
The architecture of the network is identical to that of the posterior inference network $q(\theta|t; \phi')$ used in the compression step (Section \ref{subsubsec:compressor_architecture}): the network is a MAF composed of 8 transforms, where each transform is a masked MLP with 3 layers of 256 hidden features and ReLU activation functions between the layers. 
Once trained, the normalizing flow allows both to estimate of the log-evidence $\log p(t_s)$ of a given summary statistic $t_s$ and to sample from the estimated probability distribution of the summary statistics corresponding to a given smoothing scale $k_s$. 

\section{Experiments and Results} \label{sec:results}

We now describe an example application of our method to detecting model mispecification in cosmology. In particular, we focus on detecting differences in the subgrid physics implementation across three hydrodynamical simulations and their impact on the large scale structure distribution of \emph{(i)} total matter $M_{\mathrm{tot}}$, and \emph{(ii)} gas mass $M_{\mathrm{gas}}$ distributions.

\subsection{Datasets and experimental setup} \label{subsec:datasets}
\subsubsection{CAMELS simulation suites}
Our dataset consists of 2D maps from the CAMELS project \citep{CAMELS, CAMELS_CMD, CAMELS_Astrid}. In particular, we consider total matter density $M_{\mathrm{tot}}$ and gas density $M_{\mathrm{gas}}$ maps at redshift $z = 0$ from three suites of hydrodynamical simulations: IllustrisTNG \citep{IllustrisTNG_sims_1,IllustrisTNG_sims_2}, Astrid \citep{ASTRID_sims_1, ASTRID_sims_2}, and SIMBA \citep{SIMBA_sims}, which implement different subgrid physics models. These simulations follow the evolution of 256$^3$ dark matter particles and 256$^3$ gas particles in a comoving volume of (25 $h^{-1}$ Mpc)$^3$ with periodic boundary conditions from $z=127$ to $z=0$. Each 2D map is produced by taking a $25 \times 25 \times 5 \, (h^{-1} \mathrm{Mpc})^3$ slice of the simulation box and projecting it along the third axis, so every simulation box has 15 paired maps corresponding to it. The image size of the maps is 256 $\times$ 256 pixels, which corresponds to a resolution of roughly $ \simeq 0.1 h^{-1} \mathrm{Mpc}$.

As a baseline dataset for training our models, we use the Latin Hypercube (LH) set of the Astrid suite. The simulations in the LH set vary the values of both cosmological parameters ($\Omega_M$ and $\sigma_8$) and astrophysical parameters (stellar feedback parameters, $A_{\mathrm{SN1}}$ and $A_{\mathrm{SN2}}$, and AGN feedback parameters, $A_{\mathrm{AGN1}}$ and $A_{\mathrm{AGN2}}$). This set consists of 1 000 independent simulations and 15 000 corresponding maps for each astrophysical field of interest. 

In addition to testing the parameter estimation performance of our models on a holdout test set from the Astrid LH set, we also assess the capabilities of our trained models to detect model misspecification on the Cosmic Variance (CV) sets of the three hydrodynamical simulation suites. 
The CV sets are designed to explore the effects of cosmic variance on various cosmological and astrophysical probes. For each suite, the CV set consists of 27 simulations (with 405 maps per astrophysical field) with the same values of cosmological and astrophysical parameters ($\Omega_M=0.3, \, \sigma_8=0.8, \, A_{\mathrm{SN1}}= A_{\mathrm{SN2}}= A_{\mathrm{AGN1}}= A_{\mathrm{AGN2}}=1.$), but varied random seeds for the initial conditions.  It is important to note here, however, that the astrophysical parameters do not necessarily have one-to-one correspondence between the simulation suites. We refer the readers to \cite{CAMELS} and \cite{CAMELS_Astrid} for further details on the physical meaning of the astrophysical parameters for each suite.
Examples of total matter density and gas density maps from the three CV set simulations with the same initial conditions are shown in Figure \ref{fig:astro_fields}.

To mimic a realistic observational scenario, we implement the following experimental design: we use Astrid as our baseline physics model, while treating IllustrisTNG and SIMBA as ``mock observable universes" that may be out-of-distribution relative to our baseline physics model. Specifically, we train our density estimation models on the Astrid LH set, then use the CV sets from all three suites to assess model misspecification. This approach mirrors what we would do with actual observations -- splitting the observed volume into smaller subvolumes to assess statistical significance, while all subvolumes share the same underlying physical parameters. This setup provides a controlled test of how differences in subgrid physics implementations affect our ability to detect model misspecification and obtain unbiased cosmological constraints.

\subsubsection{Implementation and training} \label{subsec:implementation_and_training}

We follow the same training procedure for both total matter density and gas density fields. We find that the similar sets of hyperparameters work well for the two astrophysical fields, although it is possible that other fields which track more distinct physical quantities, such as gas temperature or gas pressure, would require a significantly different set of hyperparameters for training.

We split the dataset for each field into training, validation, and test sets, following a random 90/5/5 split, which reserves 13 500 maps (900 distinct cosmologies) for training and 500 maps (50 cosmologies) for validation and testing. For consistency, we use the same dataset split for training both the compressor network, which is trained together with the parameter inference network, and the evidence estimation network. To aid the training, we work with the fields in the log space and standardize them prior to the training. We also apply data augmentations in the form of random rotations and mirror flips to help the models learn the relevant symmetries.

We train the compressor network using the AdamW optimizer \citep{AdamW_loshchilov2019, AdamW_kingma2017}, with peak learning rate $1 \times 10^{-4}$ for $M_\mathrm{tot}$ ($7 \times 10^{-5}$ for $M_\mathrm{gas}$) and cosine annealing learning rate schedule \cite{CosineSchedule_loshchilov2017} for 300 epochs with a batch size of 100. For the downstream task of evidence estimation we select the checkpoint with the lowest validation loss. 

The evidence estimation network is trained with similar settings: we use the AdamW optimizer with peak learning rate of $7\times 10^{-3}$ for $M_\mathrm{tot}$ ($5\times 10^{-3}$ for $M_\mathrm{gas}$) and cosine annealing scheduler to train the network for 100 epochs with a batch size of 100. Similarly to the first part of the training, we save the checkpoint with the lowest loss on the validation set.  

\begin{figure}[ht!]
    \hspace*{3cm}{{\includegraphics[height=7.5cm]{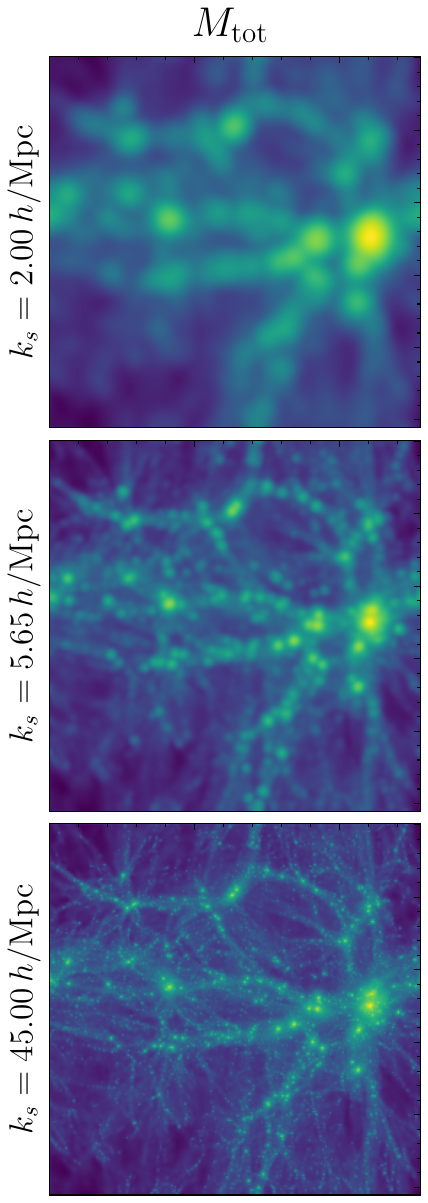} }}%
    \qquad
    {{\includegraphics[height=7cm]{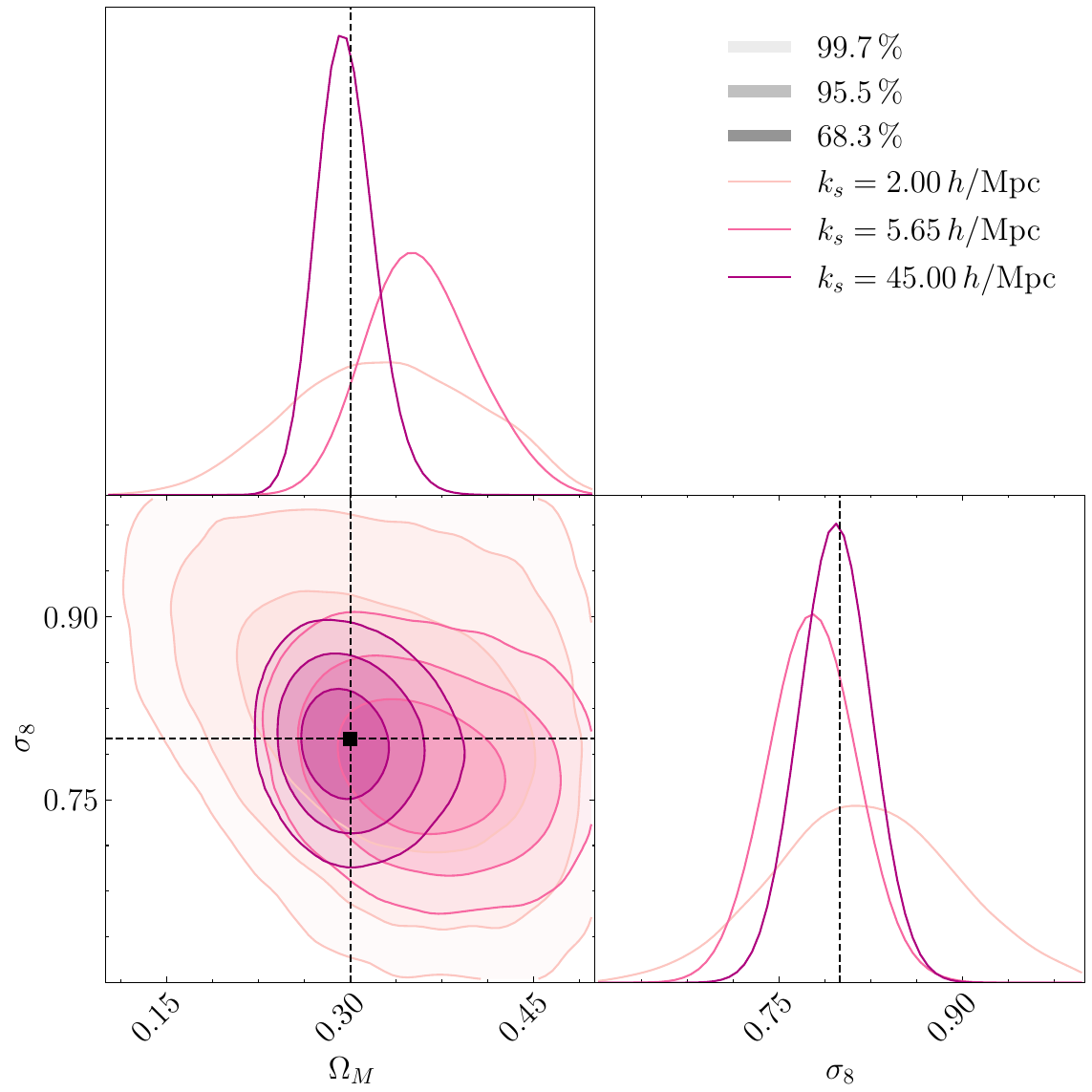} }}%

    \hspace*{3cm}{{\includegraphics[height=7.5cm]{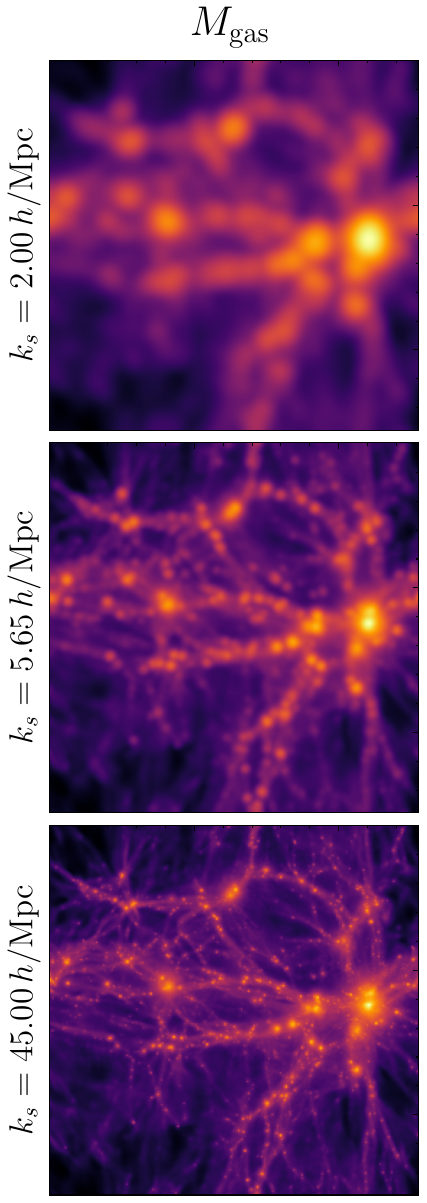} }}%
    \qquad
    {{\includegraphics[height=7cm]{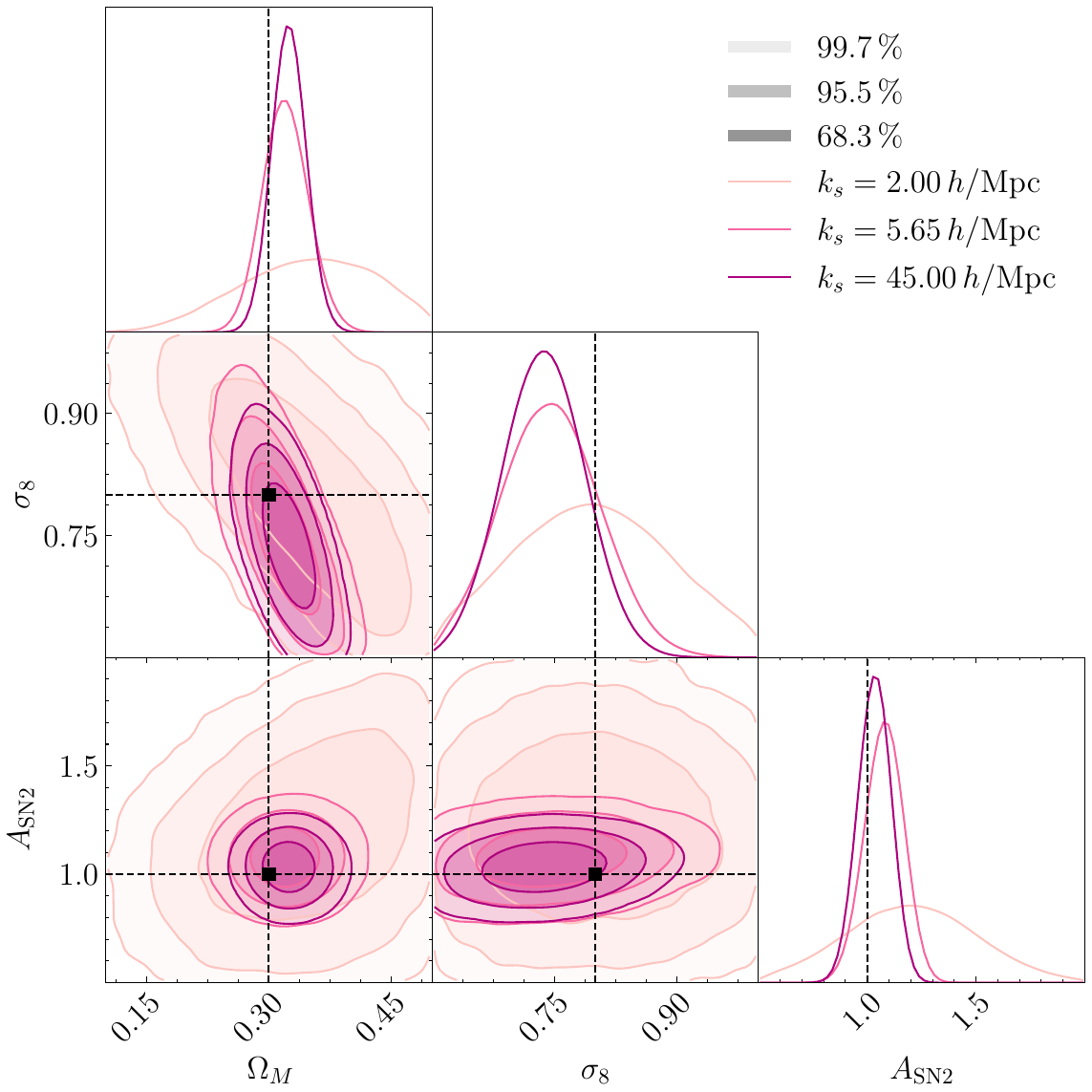} }}%
    \caption{Left: Total matter density (top) and gas density (bottom) field from the Astrid CV set with various levels of Gaussian smoothing. The smoothing scales $k_s$ plotted are 2, 5.65 and 45 $h/\mathrm{Mpc}$. Right: Corresponding inferred posteriors over astrophysical and/or cosmological parameters, with true parameter values indicated by dashed lines. Posterior contours are only plotted for the parameters for which the corresponding density fields have considerable constraining power. As we include higher frequency information, the posteriors become tighter and concentrate closer to the true parameter values, validating scale-dependence of our neural summary statistics.}
    \label{fig:npe_constraints_scale}
\end{figure}

\subsection{Validation of scale-dependent neural summary statistics} \label{subsec:neural_summary_stats}

In Figure ~\ref{fig:npe_constraints_scale}, we demonstrate how our neural summary statistics effectively capture cosmological information across different physical scales. We present parameter constraints on the cosmological parameters $\Omega_M$ and $\sigma_8$, along with relevant astrophysical parameters, as a function of smoothing scale for both total matter and gas density fields from the Astrid test set. 

As expected, incorporating higher frequency information (smaller scales) leads to significantly tighter parameter constraints for both cosmological and astrophysical parameters, confirming that our neural summarizer successfully extracts scale-dependent information. This scale-dependent behavior is crucial for understanding which physical processes dominate at different scales and for detecting model misspecification. 
In all cases, our method successfully recovers the true parameters within the expected uncertainty, validating the accuracy of our neural posterior estimation approach. 
A more thorough validation is shown in Figure~\ref{fig:coverage_all} through a coverage test as a function of smoothing scale.

\subsection{Anomaly detection with neural summaries} \label{subsec:anomaly_detection_neural_summary_stats}

\begin{figure}%
    \centering
    {{\includegraphics[height=4.5cm]{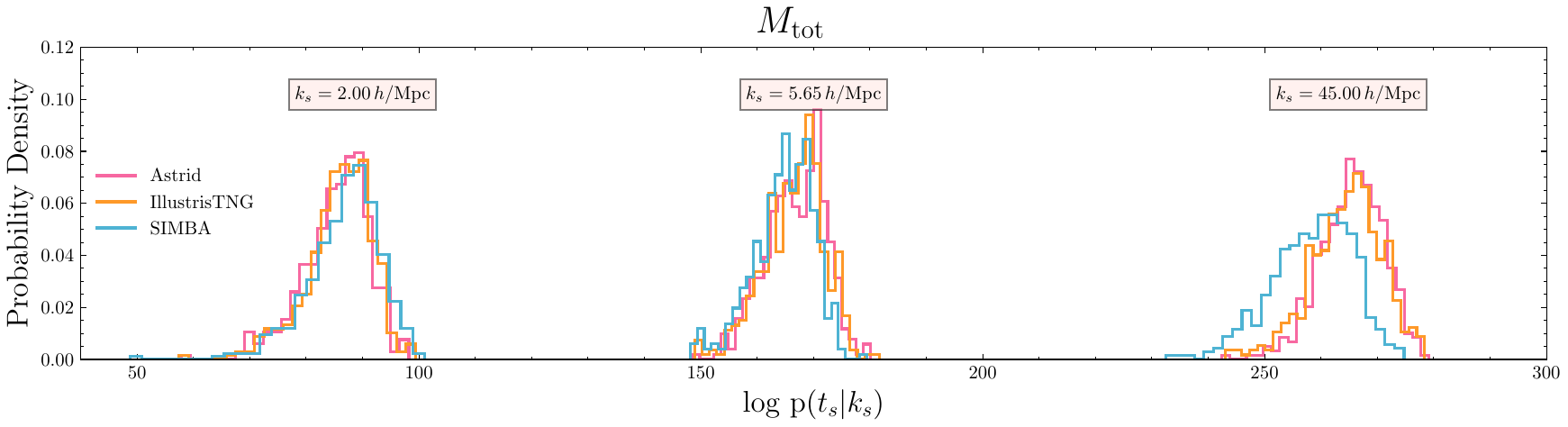} }}%
    \qquad
    {{\includegraphics[height=4.5cm]{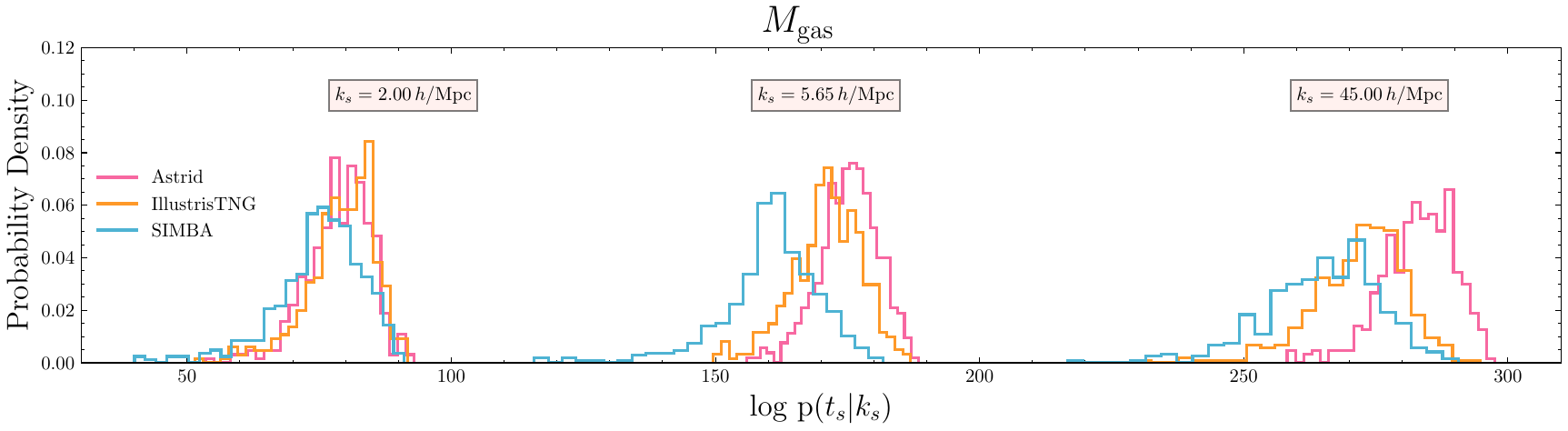} }}%
    \caption{Comparison of the distribution of log-evidence  $\log p(t_s|k_s)$ of the learned neural summary statistics for three different smoothing scales ($k_s=$ 2, 5.65, and 45 $h/\mathrm{Mpc}$) for the CV sets of Astrid, IllustrisTNG, and SIMBA suites. The top plot corresponds to the total matter density field, the bottom plot corresponds the gas density field. At large smoothing scales ($k_s = 2 h/\mathrm{Mpc}$), the distributions for the three different suites largely overlap. To visually separate the distributions for each smoothing scale, we added the following relative offsets to the distributions for $k_s = 5.65 \, h/\mathrm{Mpc}$ and $k_s = 45 \, h/\mathrm{Mpc}$: (i) 100 and 200 for $M_{\mathrm{tot}}$, and (ii) 110 and 220 for $M_{\mathrm{gas}}$. As we include more information from smaller scales, the differences between the distributions become more pronounced, in particular for the gas density fields. }
    \label{fig:hist_evidence}
\end{figure}

Having validated our neural summary statistics, we next leverage them to detect model misspecification between the simulation suites. We evaluate the log-evidence $\log p(t_s)$ of the summary statistics as a function of smoothing scale for the CV sets from the three suites using the evidence estimation network trained exclusively on the Astrid suite, our reference model. Figure~\ref{fig:hist_evidence} shows evidence distributions for three different smoothing scales. 

For total matter density fields (top row),  at both large ($k_s = 2.00 \, h/\mathrm{Mpc}$) and intermediate ($k_s = 5.65 \, h/\mathrm{Mpc}$) smoothing scales all three simulation suites produce similar evidence distributions, suggesting that the large-scale total matter distribution is relatively insensitive to differences in subgrid physics implementations. However, as we probe smaller scales ($k_s \approx 45.00 \, h/\mathrm{Mpc}$), the SIMBA distribution shifts towards lower evidence values, indicating that its small-scale physics implementation produces structures that appear increasingly ``anomalous" or out of distribution when evaluated against the Astrid model, while the IllustrisTNG distribution remains almost undistinguishable from Astrid's, related to IllustrisTNG's feedback model being more similar to that of Astrid than SIMBA's is. 

The gas density fields (bottom row) show more pronounced differences between simulation suites, as expected: unlike total matter, gas distributions also show substantial differences at intermediate scales, with both IllustrisTNG and SIMBA consistently producing lower evidence values compared to Astrid. This reflects the greater sensitivity of gas to baryonic feedback processes, which are implemented differently across the simulation suites. The differences become more extreme at smaller scales, where the complex interplay of cooling, star formation, and feedback mechanisms leads to distinct gas morphologies, as is visually evident in Figure~\ref{fig:astro_fields}.

To quantify the differences in $\log p(t_s)$ systematically, we compute the mean percentile rank of evidence values for each CV suite relative to the Astrid reference distribution. While the samples within each CV suite are not strictly independent (some of them are projections of the same simulation box), the percentile ranking provides a robust measure of relative model discrepancy. 

Figure~\ref{fig:misc_area_vs_percentile} shows the mean percentile (solid lines) for total matter and gas density fields with error bars representing the standard error on the mean percentile.
For $M_\mathrm{tot}$ (left), SIMBA's percentile rank drops dramatically from over $50\%$ at large scales to about $20\%$ at small scales, indicating increasingly significant model differences. IllustrisTNG shows a much more moderate decrease to around $45\%$, maintaining this level across intermediate and small scales. This suggests that while both models differ from Astrid, SIMBA implements substantially different small-scale physics that affects the total matter distribution. For $M_\mathrm{gas}$ (right), both simulations show considerably lower percentile ranks at small scales, dropping below $20\%$, confirming their gas distributions are highly distinct from Astrid, even at the level of the summary statistics. 
These results demonstrate our method's effectiveness at quantifying simulation differences across scales and provide valuable insights into which physical scales and components are most affected by modeling choices. This scale-dependent analysis could guide the refinement of subgrid physics models and help identify which aspects of the simulations most urgently require calibration against observational data. 

\subsection{Correlations between OOD detection and parameter bias} \label{subsec:bias_correlations}

\begin{figure}%
    \centering
    {{\includegraphics[width=8cm]{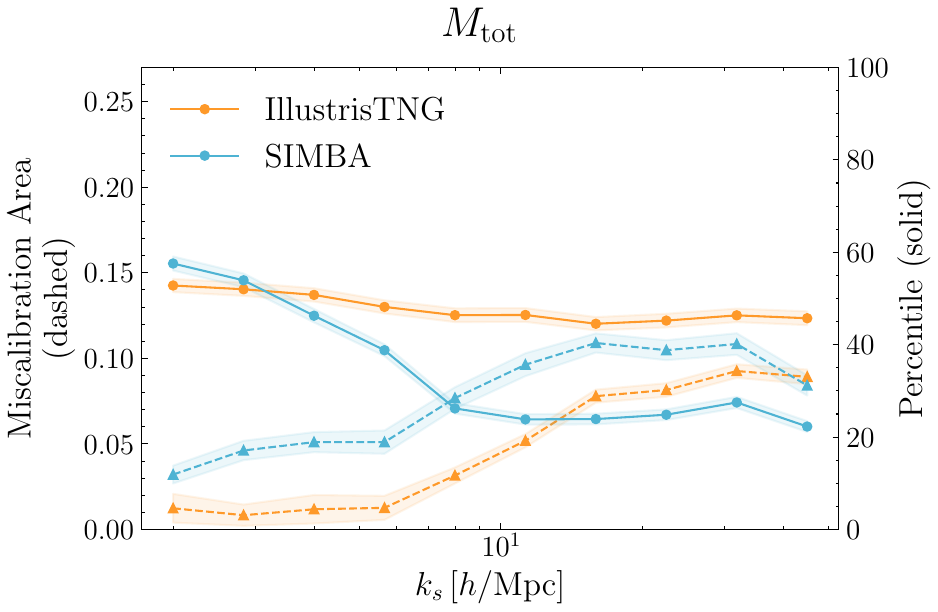} }}%
    {{\includegraphics[width=8cm]{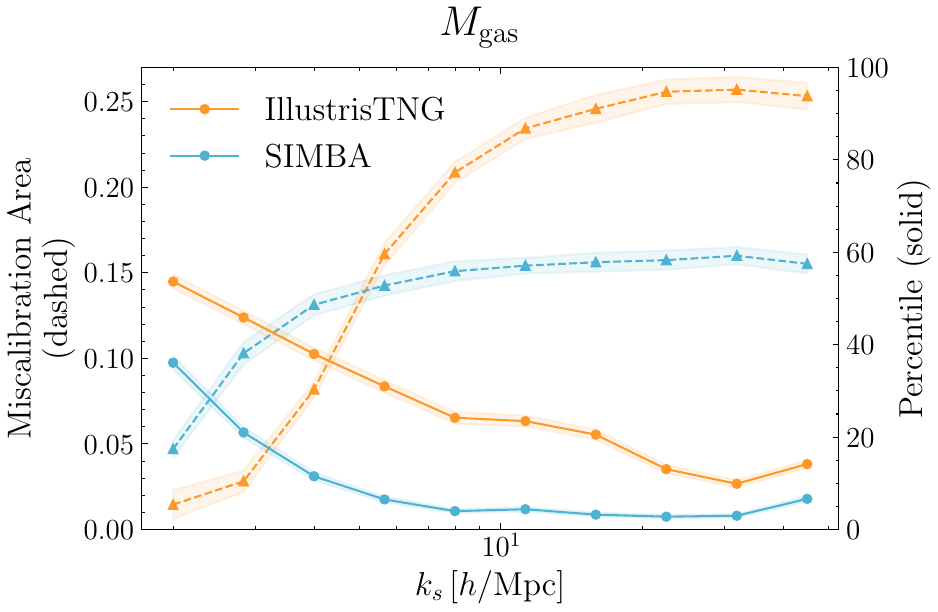} }}%
    \caption{Miscalibration area (dashed) and percentiles of the $\log p(t_s|k_s)$ values (solid) for the total matter density (left) and gas density (right) fields from of IllustrisTNG and SIMBA CV suites. The average percentile ranks are computed on the IllustrisTNG and SIMBA CV sets from these simulation suites, with respect to Astrid CV. Solid lines represent average percentile over all maps in the CV suite, while the shaded regions correspond to 1-$\sigma$ error. The miscalibration areas are computed on the test sets of IllustrisTNG and SIMBA LH suites for cosmological parameters $\Omega_M$ and $\sigma_8$. The shaded regions correspond to 1-$\sigma$ errors computed over 100 bootstrap realizations.}
    \label{fig:misc_area_vs_percentile}
\end{figure}

As discussed previously, physical degeneracies between different simulation models can complicate OOD detection. This creates two challenging scenarios: \emph{(1)} cases where model misspecification remains undetected yet produces biased parameter constraints, and \emph{(2)} cases where data appear strongly out-of-distribution while still yielding unbiased cosmological constraints. The latter occurs when the detected differences are orthogonal to the effects of variations in the cosmological parameters whose inference remains unbiased.

In Figure~\ref{fig:misc_area_vs_percentile}, we quantitatively analyze this relationship for our simulation suites between OOD detection and parameter bias across our simulation suites. We use the average percentile rank as our OOD metric. To quantitatively assess parameter bias from the inferred posteriors, we looked for a metric which \emph{(i)} assesses parameter bias in 2D space over both cosmological parameters, \emph{(ii)} accounts for correlated biases (such as a constant offset in predicted parameters), and \emph{(iii)} penalizes models more for overconfidence and correlated biases than for underconfidence. To satisfy the above criteria, we choose to use `miscalibration area' as our bias measure.
We plot both the average percentile rank (our OOD metric) and the miscalibration area (our bias measure) as functions of smoothing scale.

To compute the miscalibration area, we first compute the expected coverage for cosmological parameters on the test datasets of IllustrisTNG and Astrid LH sets. We do not compute the coverage for the astrophysical parameters since these parameters do not represent the same physical quantities for the different subgrid models.  
Similarly to~\cite{misc_area_tran2020methods}, we define the miscalibration area as the area between the expected coverage curve and the diagonal line between 0 and 1 (which represents perfectly calibrated model). Defined this way, the miscalibration area spans the range of values [0, 0.5], with lower values correlating with smaller parameter bias. We provide further details on the choice of miscalibration area as our bias metric and its computation for under- and over-confident posteriors in Appendix~\ref{appendix:misc_area}.

For both total matter and gas density fields, we observe a notable correlation between the OOD detection strength and parameter bias. As the percentile ranks drop to lower values on smaller smoothing scales, the miscalibration area increases appreciably from less 0.05 at large $k_s$ to around 0.1 for $M_{\mathrm{tot}}$ and 0.15-0.25 for $M_{\mathrm{gas}}$. These correlations suggest that that the differences in baryonic physics between simulation suites directly impact the summary statistics most relevant to cosmological parameter inference (top right quadrant in Figure~\ref{fig:quadrant}), making the observables related to matter, such as lensing, and gas density fields, such as tSZ and kSZ, a potentially reliable probe for detecting physically relevant model misspecification.

We note, however, that for $M_\mathrm{gas}$ fields (right panel), the parameter bias in the IllustrisTNG simulations appears to be considerably higher than the parameter bias in the SIMBA simulations, although their higher percentile rank would indicate that they are more in-distribution with respect to the baseline Astrid simulations. This feature underscores the above mentioned limitation: although it could help estimate relative increase in parameter bias with the inclusion of small-scale information, OOD detection strength alone cannot reliably predict the magnitude of parameter bias. This highlights the importance of understanding the specific physical degeneracies between cosmological parameters and astrophysical effects when interpreting both OOD metrics and parameter constraints. 

\section{Summary and Discussion} \label{sec:summary}

In this paper, we have presented a general framework for detecting model misspecification in cosmological datasets as a function of scale. 
Our framework uses Bayesian evidence of neural summary statistics as a metric to assess the degree of the model misspecification between the baseline training data and observations. 
As part of our framework, we propose a training strategy for constructing neural summary statistics of cosmological fields by training a single neural network model which is conditioned on the smoothing scale applied to the fields. 
These summary statistics are learned from the data to extract all relevant information about the parameters of interest using the VMIM approach. 
The learned statistics can then be used for a couple of complementary purposes, which include (1) obtaining constraints on the parameters of interest, and (2) as an informative low-dimensional representation of the data for Bayesian evidence estimation.

Using total matter density and gas density fields from CAMELS simulations as a first application, we demonstrate that this framework allows us to detect the discrepancies in the subgrid physics between different simulations suites (IllustrisTNG, Astrid, SIMBA). 
In particular, for both types of fields, we find that, as the fields capture more information from small scales, simulated fields from IllustrisTNG and SIMBA suites become increasingly easier to identify as being out-of-distribution with respect to our baseline dataset (Astrid). 
This is consistent with our physical intuition, as the differences in the subgrid physical models affect matter and gas distribution on small scales to a larger extent. 

We find the OOD detection strength to be correlated with the bias on inferred cosmological parameters for both $M_\mathrm{tot}$ and $M_\mathrm{gas}$ fields. However, as noted previously, there exists a fundamental degeneracy between the OOD detectability and parameter bias (Figure \ref{fig:quadrant}) which implies that OOD detectability does not necessarily correspond to biased parameter inference. 
This limitation is shared by OOD detection methods \textit{in general}, and breaking this degeneracy in future OOD detection analyses can be approached by, for example, using multiple complementary probes.

Our proposed framework enables us to identify model misspecification in the representation space of neural summary statistics of the simulated fields. 
These summary statistics are learned via the VMIM approach designed to preserve all relevant information for inferring parameters of interest. 
As a future direction of this work, it would be interesting to investigate how the OOD detectability changes with a different choice of representation space. 
For example, among others, self-supervised learning (SSL) methods offer alternative ways of learning neural summary statistics of the data and, as a result might, capture aspects of the data in the representations (see \cite{gui2024_ssl_survey} for a review of SSL methods and \cite{Hayat_2021_ssl}, \cite{huertascompany2023_ssl_review}, \cite{Akhmetzhanova_2023_ssl}, \cite{parker2024_ssl_astroclip} for examples of applications of these methods to astrophysical datasets). 
Another promising avenue for exploration is assessing the model misspecification at the field-level by estimating the evidence using deep generative models which provide access to the exact likelihood of the data, such as diffusion models (\cite{cuesta2024point}, \cite{nayantara_mudur2024diffusion}) and more complex normalizing flows models for modeling images(\cite{friedman2022_higlow}). 
    
We use the density fields from the CAMELS simulations as a proof-of-concept training dataset. A practical application of this framework to realistic cosmological observations, such as, for example, observations from galaxy surveys such as DESI, would necessitate a number of extensions both in terms of the methodology and of the complexity of the training data, outlined below. 

The CAMELS simulations span a volume of $(25 h^{-1}/\mathrm{Mpc})^3$, which is much smaller than the effective volumes of astrophysical surveys of interest. One important direction for future work would be to develop suites of large scale simulations which directly model observed tracers and incorporate realistic survey systematics, such as
selection biases, redshift-space distortions, and fiber collisions (similarly to, for example, \cite{lemos2023_SimBIG}). 

The distribution of galaxies from surveys such as DESI is most naturally represented as 3-D point cloud -- a collection of points in 3-D space with a certain number of informative attributes (e.g. velocities) attached to each point. From the methodological point of view, it would be valuable to perform the model misspecification analysis either in the compressed latent space of the point clouds or at the field-level by working with, for instance, diffusion-based generative models that model observed galaxies as point clouds, as was previously done in \cite{cuesta2024point} and \cite{ nguyen2024dreams}. 

These additional follow-up studies are necessary to effectively extend this framework to real cosmological survey data and to make steps towards improved, more robust modeling of our universe. 

\begin{acknowledgments}
We would like to thank Francisco Villaescusa-Navarro for helpful discussions.
AA thanks the LSST-DA Data Science Fellowship Program, which is funded by LSST-DA, the Brinson Foundation, the WoodNext Foundation, and the Research Corporation for Science Advancement Foundation; her participation in the program has benefited this work.
This work is supported by the National Science
Foundation under Cooperative Agreement PHY-2019786
(The NSF AI Institute for Artificial Intelligence and
Fundamental Interactions). 
This material is based upon
work supported by the U.S. Department of Energy, Office
of Science, Office of High Energy Physics of U.S.
Department of Energy under Grant No. DE-SC0012567.
This work was performed in part at the Aspen Center for Physics, which is supported by National Science Foundation grant PHY-2210452.
The computations in this paper were run on the FASRC
Cannon cluster supported by the FAS Division of Science
Research Computing Group at Harvard University. 
\end{acknowledgments}

\software{
The code used to reproduce the results of this paper is available at \url{https://github.com/AizhanaAkhmet/model-misspecification}. 
Jupyter \citep{Kluyver2016JupyterN}, Matplotlib \citep{Hunter:2007},
Numpy \citep{harris2020array},
PyTorch \citep{NEURIPS2019_9015},
PyTorch-Lightning \citep{Falcon_PyTorch_Lightning_2019},
Scipy \citep{2020SciPy-NMeth}, and wandb \citep{wandb}.
}

\bibliography{refs}{}
\bibliographystyle{aasjournal}

\newpage
\appendix

\section{Parameter inference with total matter and gas mass density fields} \label{appendix:param_inference_no_smoothing}

\begin{figure}[ht!]
\plotone{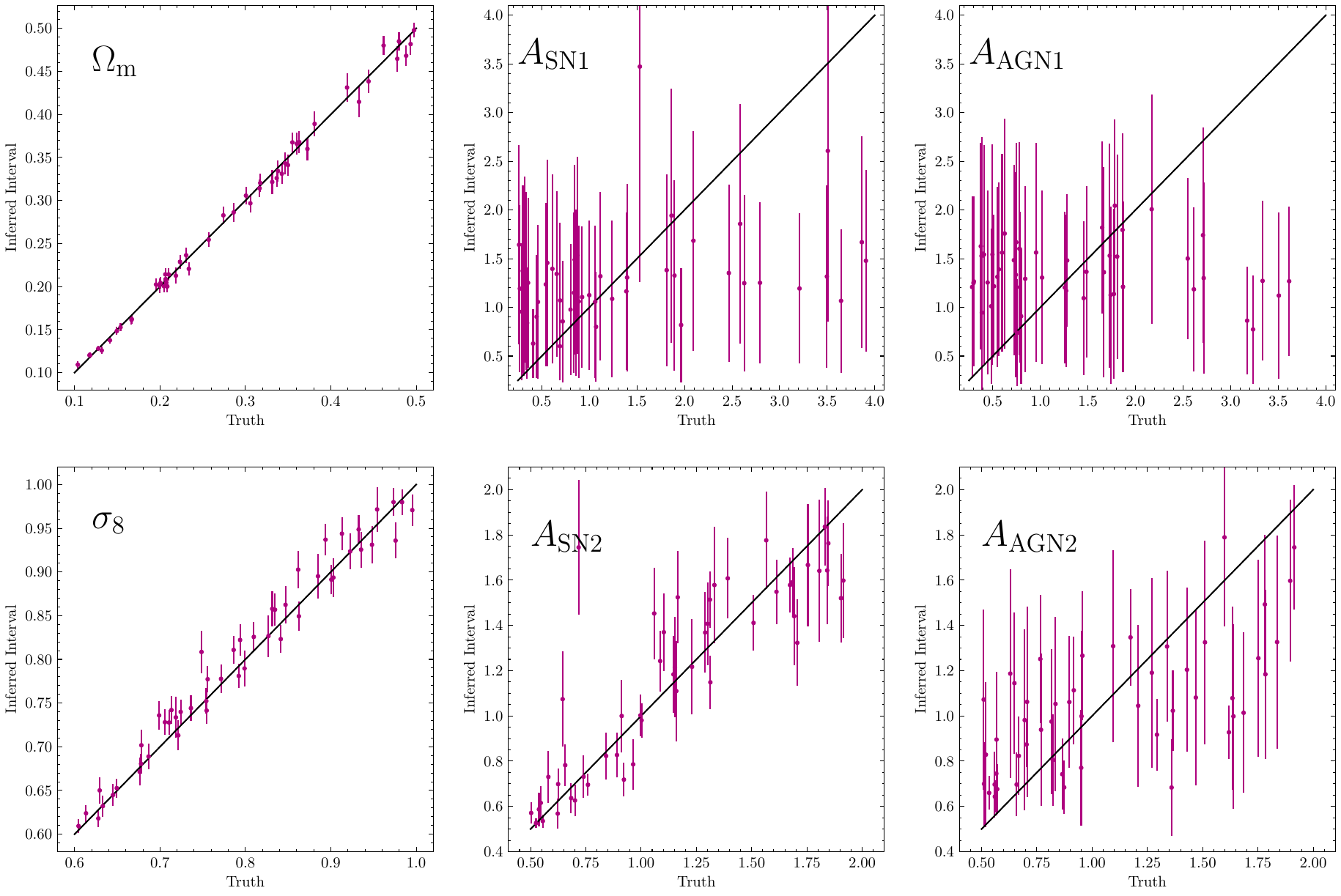}
\caption{Predicted parameters against ground-truth parameters for total matter density fields (without Gaussian smoothing) of the LH test set of the Astrid suite. The mean values and $1-\sigma$ uncertainties are computed over 10 000 posterior samples for each input field. The parameter inference network is able to constrain cosmological parameters to a high degree of both precision and accuracy.}
\label{fig:npe_Mtot_no_smoothing}
\end{figure}

\begin{figure}[ht!]
\plotone{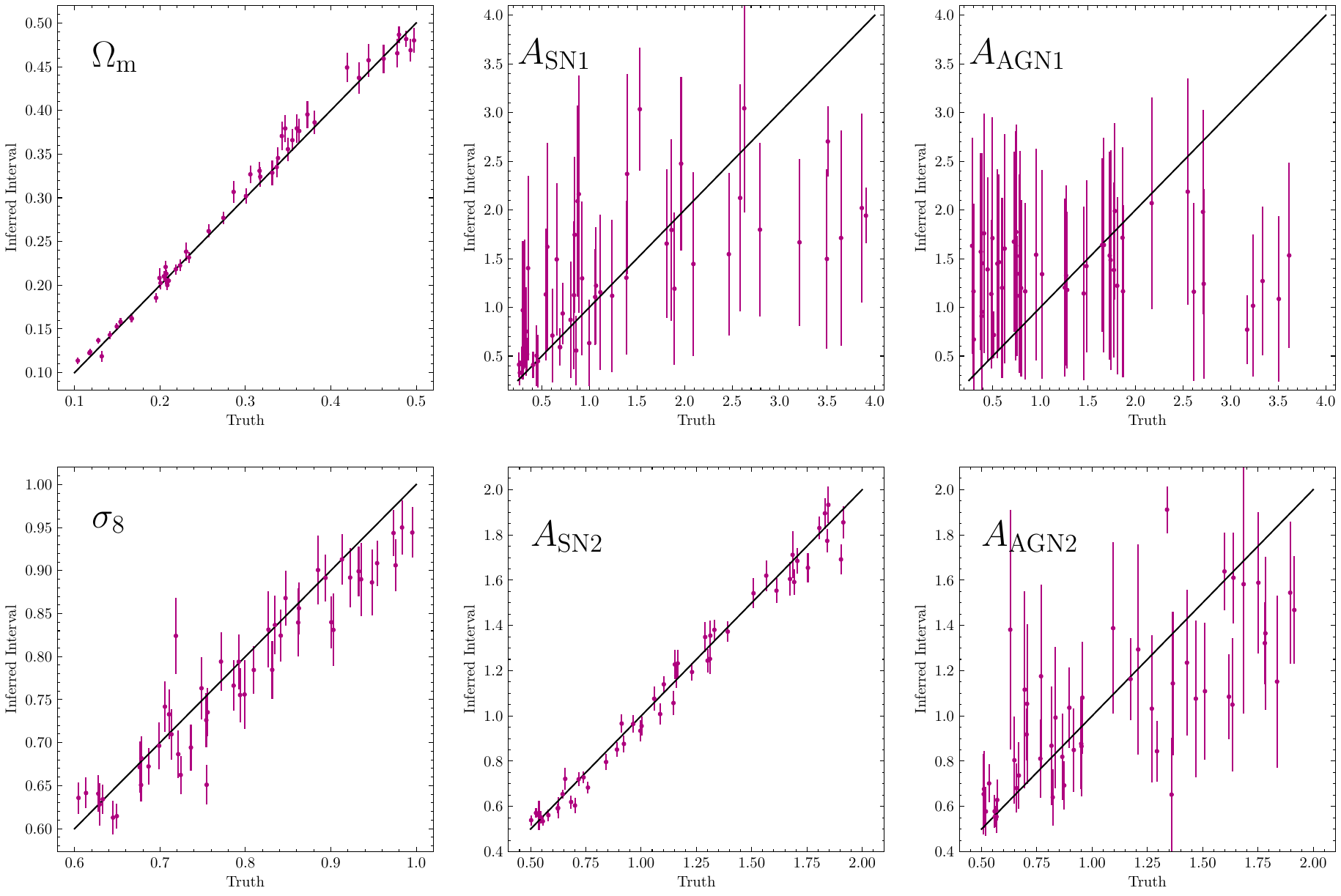}
\caption{Predicted parameters against ground-truth parameters for gas density fields (without Gaussian smoothing) of the LH set of the Astrid suite. The mean values and $1-\sigma$ uncertainties are computed over 10 000 samples for each input field. In addition to the cosmological parameters, gas density fields from Astrid can also place tight constraints on the supernovae feedback parameter $A_\mathrm{SN2}$.}
\label{fig:npe_Mgas_no_smoothing}
\end{figure}

In Figures~\ref{fig:npe_Mtot_no_smoothing} and \ref{fig:npe_Mgas_no_smoothing} we plot mean values and $1-\sigma$ uncertainties for cosmological ($\Omega_M, \, \sigma_8$) and astrophysical parameters ($A_\mathrm{SN1}, \, A_\mathrm{SN2}, \, A_\mathrm{AGN1}, \, A_\mathrm{AGN2}$) inferred from the total matter density and gas density maps from the test set of Astrid LH without Gaussian smoothing. 

We estimate the parameters from the fields following the same approach and using the same model architectures as in Section~\ref{sec:methodology}, but without conditioning the summary statistics on the smoothing scale $k_s$. We use the same training/validation/test split and training strategy as in Section~\ref{subsec:implementation_and_training}, with the peak learning rate of $7 \times 10^{-5}$ for both fields. 

We find that matter density fields can primarily constrain the cosmological parameters, but have considerably less predictive power for the astrophysical parameters, while using the gas density field one can also place tight constraints on the supernovae feedback parameter $A_\mathrm{SN2}$. Therefore, when validating our compressor and posterior inference networks in Appendix~\ref{appendix:param_inference_no_smoothing}, we estimate coverage for cosmological parameters only for $M_{\mathrm{tot}}$ fields and for cosmological parameters and supernovae feedback parameter $A_\mathrm{SN2}$  for $M_{\mathrm{gas}}$ fields.

\section{TARP Coverage Tests} \label{app:TARP}
\begin{figure}
    \centering
    {{\includegraphics[width=\textwidth]{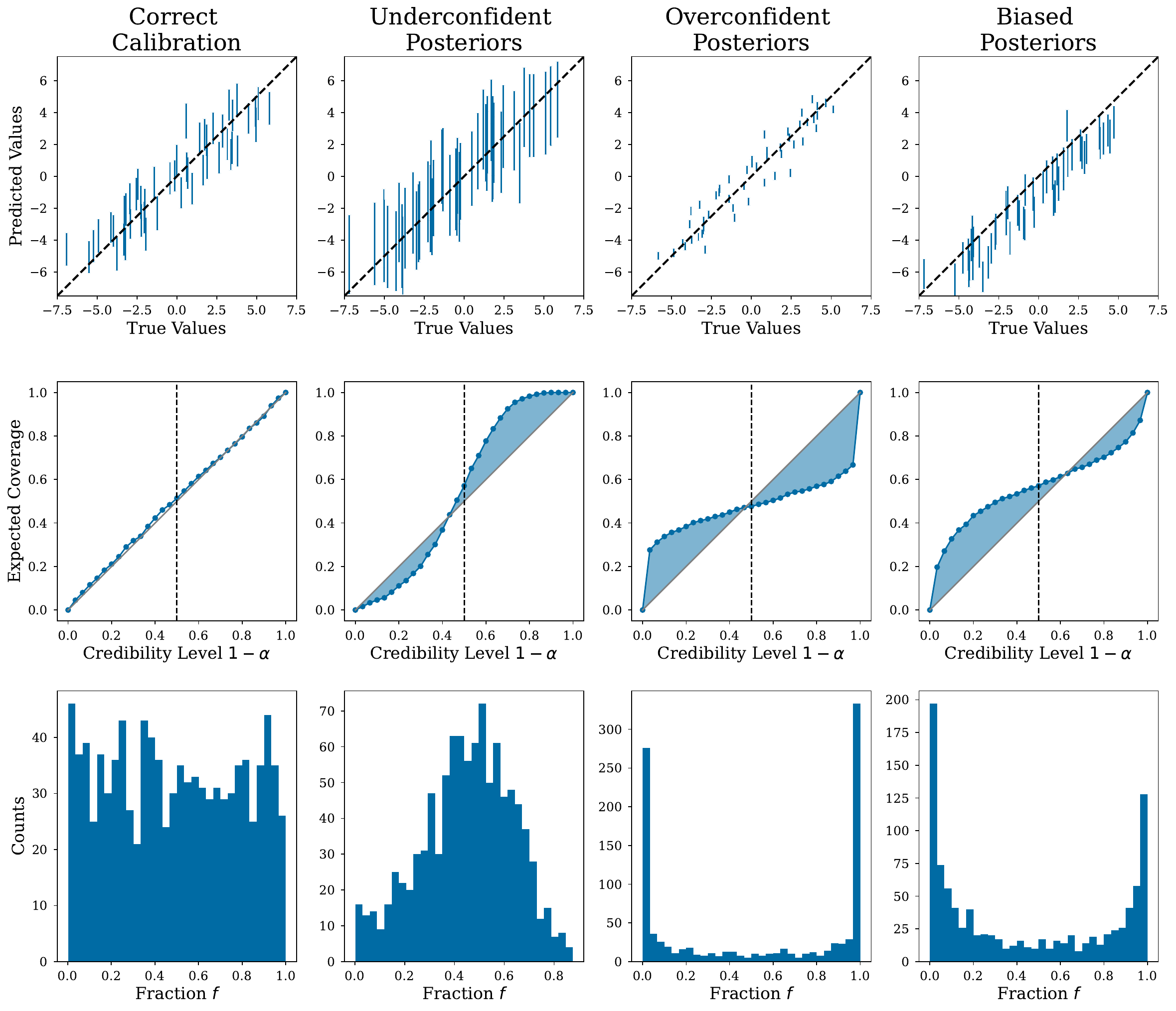} }}
    \caption{Examples of TARP coverage plots for a toy 1-D Gaussian model for (i) well-calibrated posteriors, (ii) underconfident posteriors, (iii) overconfident posterios, and (iv) biased posteriors. Top row: predicted mean values and $1-\sigma$ errors plotted against the true values. Middle row: Expected coverage probabilities (ECP) as function credibility level $1 - \alpha$. Bottom row: Histogram of fractions $f$ of posterior samples within TARP credible regions. ECPs and histograms of fractions $f$ look qualitatively different for the four cases, with overconfident and biased posteriors displaying most similarity.}
    \label{fig:toy_example_TARP}
\end{figure}
We use the TARP method \citep{lemos2023tarp} to assess the calibration of our inferred posteriors. Given a distance metric $d: U \times U \rightarrow R$ and a reference parameter point $\theta_r$, a TARP credible region for a parameter-data pair $(\theta^\star, x^\star)$ and a posterior estimator $q(\theta|x)$ is defined as follows:
\begin{equation}
    \mathcal{D}_{\theta_r} := \{ \theta \in U \, | \, d(\theta, \theta_r) \leq d(\theta^\star, \theta_r)\},
\end{equation}

A TARP confidence level $1 - \alpha$ is then defined as the integral of the estimated posterior $q(\theta|x)$ over that credible region:
\begin{equation}
 \int_{\mathcal{D}_{\theta_r} (q, \alpha, x, d)}\mathrm{d} \theta \, q (\theta| x) = 1 - \alpha. 
\end{equation}

The expected coverage probability (ECP) of the $1 - \alpha$ TARP regions is defined as
\begin{equation}
\mathrm{ECP}\left(q, \alpha, \mathcal{D}_{\theta_r}\right)=\mathbb{E}_{p(\theta, x)}\left[ \mathbbm{1} (\theta \in \mathcal{D}_{\theta_r} (q, \alpha, x^\star, d))\right]. %
\end{equation}

Empirically, ECP can be computed as the fraction of simulations of $\{\theta_i^\star, x_i\}$ for which, at a given confidence level $1 - \alpha$, a fraction $f_i$ of $n$ randomly drawn posterior samples $\{\theta_{ij}\}_{j=1}^{n}$ are a shorter distance $d(\theta_{ij}, \theta_r)$ away from a reference point $\theta_r$ than the true parameter $\theta^\star_i$:
\begin{equation}
\mathrm{ECP}\left(\hat{p}, \alpha, \mathcal{D}_{\theta_r}\right)=\frac{1}{N_{\text {sims }}}\sum_{i=1}^{N_{\text {sims }}} \mathbbm{1}\left(f_i<1-\alpha\right),
\end{equation}
with $f_i=\frac{1}{n} \sum_{j=1}^n \mathbbm{1}\left[d\left(\theta_{i j}, \theta_r\right)<d\left(\theta_i^*, \theta_r\right)\right]$.

The details of algorithm for computing the TARP coverage are presented in full in Algorithm 2 in \cite{lemos2023tarp}, and an illustration of the proposed coverage test in Figure 1 therein. 

To provide some intuition for interpreting our posterior calibration test results, we show results from a simple 1-D Gaussian toy model for four different scenarios: (i) correct calibration, (ii) underconfident posteriors, (iii) overconfident posteriors, and (iv) biased posteriors.  

In each case, the true parameters $\theta^\star$ are drawn from a normal distribution: $\theta_i^\star \sim \mathcal{N}(\bar{\theta}_i, \sigma^2)$, where $\bar{\theta}_i \in [-5, 5]$ and $\sigma^2 = 1$. The posterior samples in each scenario are drawn from the following distributions:
\begin{enumerate}[(i)]
    \item Correct calibration: $\theta_{ij} \sim \mathcal{N}(\bar{\theta}_i, \, \sigma^2)$,
    \item Underconfident posteriors: $\theta_{ij} \sim \mathcal{N}(\bar{\theta}_i, \, 2.5 \, \sigma^2)$,
    \item Overconfident posteriors: $\theta_{ij} \sim \mathcal{N}(\bar{\theta}_i, \, 0.25 \, \sigma^2)$,
    \item Biased posteriors: $\theta_{ij} \sim \mathcal{N}(\bar{\theta}_i - 1.5, \, \sigma^2)$.
\end{enumerate}

We plot the predicted mean values and 1-$\sigma$ errors against the true values $\theta^\star_i$ for all cases in the top row of Figure~\ref{fig:toy_example_TARP}. 

The second and third rows of Figure~\ref{fig:toy_example_TARP} show the corresponding expected coverage probabilities (ECP) and histogram of fractions $f_i$ used to compute the ECPs  for the four scenarios. As can be seen from the Figure, for well-calibrated posteriors, the expected coverage should match the corresponding the credible level for all credible levels $(1 -\alpha) \in [0, 1]$. For miscalibrated posteriors, the ECP curves deviate from the diagonal in a characteristic manner, which depends on the type of miscalibration. 

Note that the TARP coverage plots are different from coverage plots obtained with another well-known method, the highest-posterior density (HPD) estimation. The histograms on the bottom row of Figure~\ref{fig:toy_example_TARP} can provide some intuition behind these difference. 
\begin{itemize}
    \item For correctly calibrated posteriors, the fractions of points $f_i$ within a radius $d(\theta_i^\star, \theta_r)$ is distributed roughly uniformly, so the ECP curve, which is proportional to the cumulative sum of $f_i$, matches the credible levels $(1 -\alpha)$. 
    \item For underconfident case, since the posteriors are quite wide, . 
    \item For the overconfident posteriors, the probability mass is narrowly concentrated around the predicted mean value, for a randomly chosen reference point $\theta_r$, it is likely that either there would be very few posterior samples $\theta_{ij}$ which are closer to $\theta_r$ than $\theta^\star_i$ ($f_i \simeq 0$), or most of the posterior samples would be closer to $\theta_r$ than $\theta^\star_i$ ($f_i \simeq 1$). We see this bi-modal behavior on the histogram in the bottom row of Figure 4, which in turn results in the characteristic shape of the ECP curve.
    \item The biased posteriors somewhat mirror the overconfident posteriors: since the truth $\theta_i^\star$ is always far from the peak of the posterior, TARP credible regions are likely to either cover most of the posterior distribution $f_i \simeq 1$ or almost none of it ($f_i \simeq 0$). 
\end{itemize}

Appendix B in ~\cite{lemos2023tarp} provides additional examples to gain some intuition about these differences. 

The shaded regions on Figure~\ref{fig:toy_example_TARP} show the `miscalibration area' we use in Sec.~\ref{subsec:bias_correlations} to assess the magnitude of bias in the inferred posteriors. Since we do not want to overly penalize underconfident posteriors (compared to overconfident or biased posteriors), we compute the miscalibration area as follows:
\begin{itemize}
    \item For overconfident and biased posteriors (as indicated by the coverage plots), miscalibration area is computed as the total area between the ECP and the diagonal line which corresponds to the perfect calibration.
    \item For underconfident posteriors (as indicated by the coverage plots), miscalibration area below is the diagonal is counted as 'negative', while the area above the diagonal is taken to be 'positive'. The total miscalibration area is then the absolute value of the sum of the two.
\end{itemize}

\section{Posterior Calibration Test} \label{appendix:posterior_calibration}

We now examine whether the posterior inference networks for $M_\mathrm{tot}$ and $M_\mathrm{gas}$ fields trained on Astrid are well-calibrated with the TARP method \citep{lemos2023tarp}. We use \texttt{tarp} package with Euclidean distance metric to evaluate the expected coverage for our posteriors. 

We plot the expected coverage for the two fields in the first column Figure~\ref{fig:coverage_all}. The coverage is computed as an average coverage across multiple parameter-data pairs on the test set of Astrid LH, and the error bars are $1-\sigma$ errors computed over 100 bootstrap realizations. As noted in App.~\ref{appendix:param_inference_no_smoothing}, we compute the coverage only for the parameters for which the density fields have constraining power for. 
We compute the expected coverage for a wide range of smoothing scales which are logarithmically spaced from $k_{s, \mathrm{min}} = 2\, h/\mathrm{Mpc}$ to $k_{s, \mathrm{max}}=45 \, h/\mathrm{Mpc}$ and find the posteriors to be fairly well-calibrated for all of the scales within this range.

\section{Miscalibration Area from TARP coverage tests} \label{appendix:misc_area}
\begin{figure}
    \centering
    {{\includegraphics[width=\textwidth]{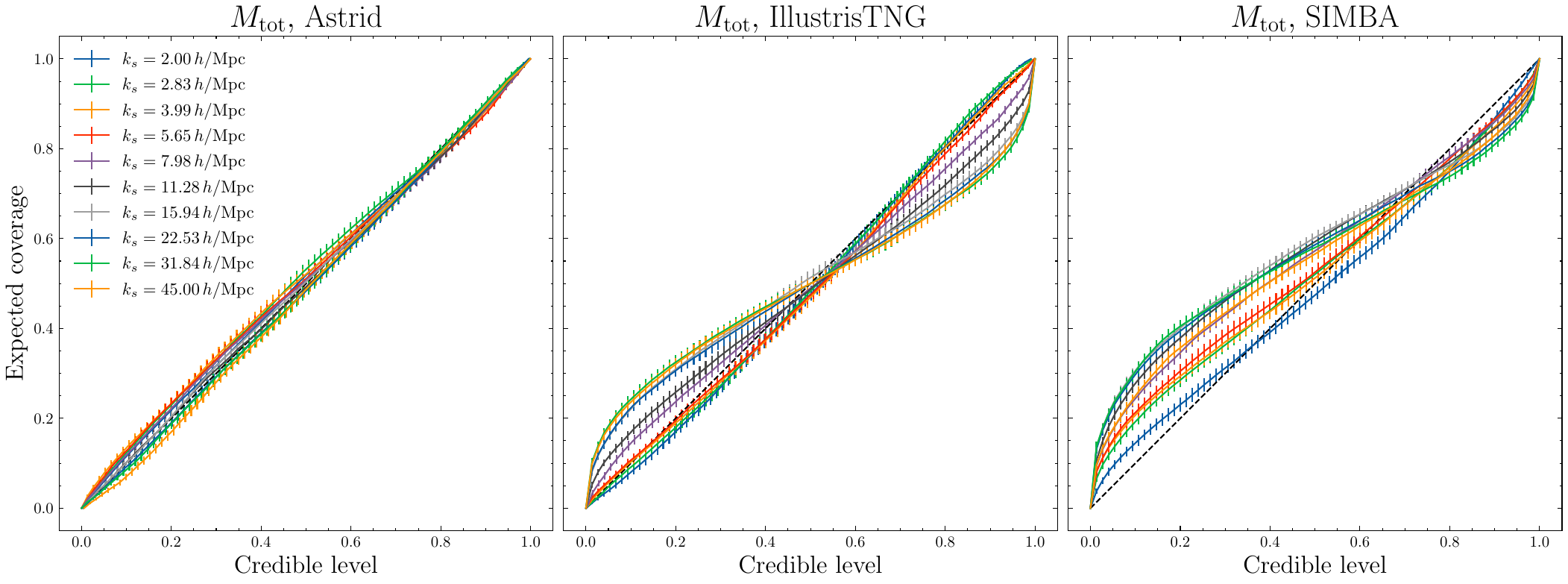} }}
    {{\includegraphics[width=\textwidth]{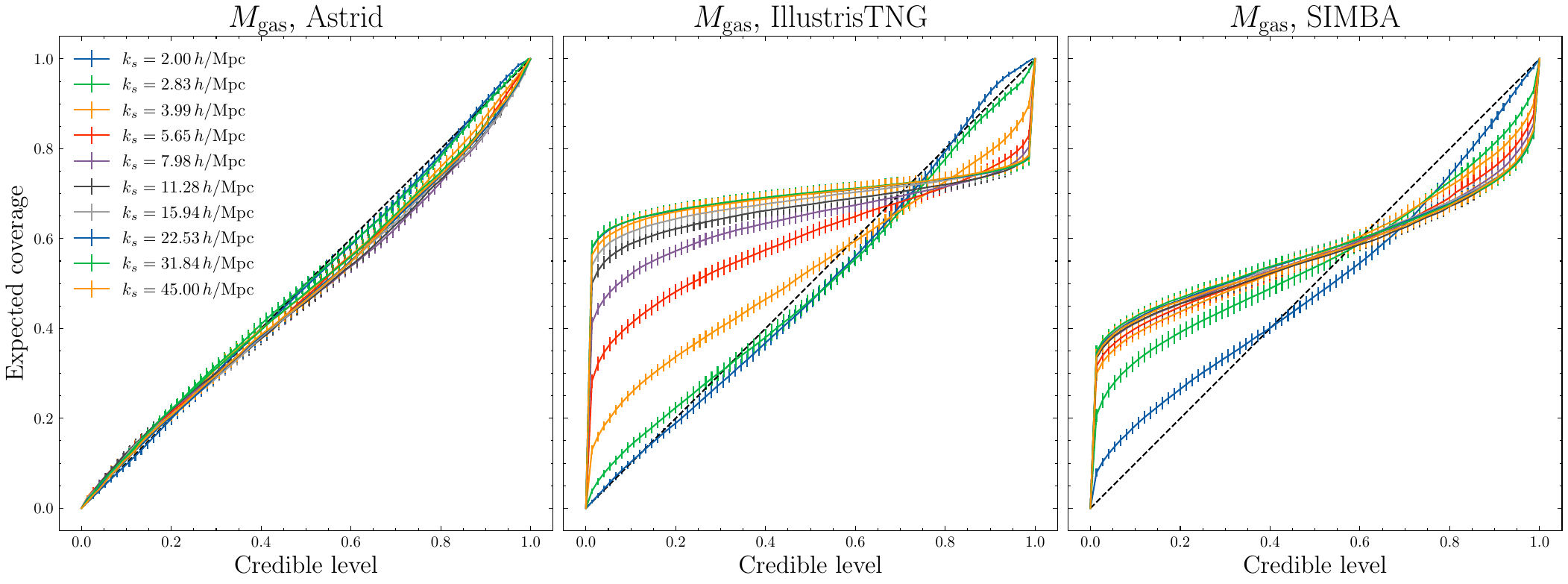} }}
    \caption{Expected coverage versus confidence levels computed with the TARP method \citep{lemos2023tarp} for a range of smoothing scales $k_s$ for $M_\mathrm{tot}$ (top) and $M_\mathrm{gas}$ fields. The error bars are $1-\sigma$ errors computed over 100 bootstrap realizations. The posterior inference networks are well-calibrated for Astrid (used as the in-distribution training dataset) on all smoothing scales, but are poorly calibrated (mostly overconfident and biased) when tested on the IllustrisTNG and SIMBA suites (out-of-distribution datasets).}
    \label{fig:coverage_all}
\end{figure}

We compute expected coverages using TARP method for IllustrisTNG and SIMBA using with the compressor and posterior inference models trained on Astrid. In Figure~\ref{fig:coverage_all}, we show the coverage plots for the two simulation suites for the same range of smoothing scales as in App.~\ref{appendix:posterior_calibration}. The error bars are $1-\sigma$ errors computed over 100 bootstrap realizations. For $M_\mathrm{tot}$ and $M_\mathrm{gas}$ from both simulations, the posteriors become increasingly miscalibrated on smaller smoothing scales (higher $k_s$). 

The characteristic shapes of the expected coverage curves indicate that the posteriors are primarily overconfident and biased, except for IllustrisTNG simulations on large smoothing scales, and we compute the corresponding miscalibration areas accordingly (see App.~\ref{app:TARP} for details). 

\end{document}